\documentclass[12pt]{article}

\usepackage[dvips]{graphicx}

\def\dspace{\baselineskip = 0.30in}

\def\lapproxeq{\lower .7ex\hbox{$\;\stackrel{\textstyle
<}{\sim}\;$}}
\def\gapproxeq{\lower .7ex\hbox{$\;\stackrel{\textstyle
>}{\sim}\;$}}

\begin{document}

\dspace

\begin{titlepage}
\begin{flushright}
BA-03-06\\
SNUTP 03-011
\end{flushright}
\vskip 2cm
\begin{center}
{\Large\bf Brane gravity, massless bulk scalar and
\\
~self-tuning of the cosmological constant 
}
\vskip 1cm {\normalsize\bf
$^{(a)}$Jihn E. Kim\footnote{jekim@phyp.snu.ac.kr},
$^{(b)}$Bumseok Kyae\footnote{bkyae@bartol.udel.edu}, and
$^{(b)}$Qaisar Shafi\footnote{shafi@bartol.udel.edu}
}
\vskip 0.5cm
{\it
$^{(a)}$School of Physics, Seoul National University, Seoul 151-747, Korea
\\
$^{(b)}$Bartol Research Institute, University of Delaware, Newark,
DE 19716, USA\\[0.1truecm]}

\end{center}
\vskip 0.8cm

\begin{abstract}

We show that a self-tuning mechanism of the
cosmological constant could work in 5D non-compact space-time with
a $Z_2$ symmetry in the presence of a massless scalar field. 
The standard model matter fields live only on the
4D brane. The change of vacuum energy on the brane
(brane cosmological constant) by, for instance, electroweak and
QCD phase transitions, just gives rise to dynamical shifts of the
profiles of the background metric and the scalar field in the extra
dimension, keeping 4D space-time flat without any fine-tuning. To
avoid naked singularities in the bulk, the brane cosmological
constant should be negative. We introduce an additional
brane-localized 4D Einstein-Hilbert term so as to provide the
observed 4D gravity with the non-compact extra dimension. 
With a general form of brane-localized gravity term 
allowed by the symmetries,    
the low energy Einstein gravity is successfully reproduced on the brane 
at long distances.  
We show this phenomenon explicitly for the case of vanishing bulk
cosmological constant.
\end{abstract}

\end{titlepage}

\newpage


\section{Introduction}

The problem of the cosmological constant or the absolute scale of
the potential energy has been considered as one of the most
important and difficult problems in modern particle
physics~\cite{weinberg}. In recent years, however, there has been a
hope to understand the vanishing cosmological constant in extra
dimensional field theories~\cite{kachru,nilles,kkl1}. 
In four dimensional (4D) space-time, the vacuum energy $V_0$ ($=3M_P^2R$)
determines the curvature of our space-time. The flat 4D space-time is
possible only for $V_0=0$. Thus, in 4D one must fine-tune the
vacuum energy to zero to obtain a flat space-time.  However, it is
more involved in higher dimensional field theories.

In the Randall-Sundrum-II (RS-II) type five dimensional (5D) theories
with a 3-brane embedded in a {\it non-compact} 5D space,
the 5D gravity sector contains a 5D cosmological constant $\Lambda_b$ and
a brane vacuum energy $\Lambda_1$~\cite{rs2}.
For a flat 4D subspace, a fine-tuning relation
between $\Lambda_b$ and $\Lambda_1$
is inevitable in the simple RS-II model~\cite{rs2}.
However, by introducing a massless scalar field in the bulk,
it was shown that a self-tuning solution of the cosmological constant
is possible in the RS-II type setup~\cite{kkl1}.
To guarantee a massless scalar field in 5D, the four-form field
strength $H_{MNPQ}$ has been very useful~\cite{kkl1,townsend}.

In the initial attempts of RS-II type models, one restricted to a
finite value for the integration of the warp factor
through the extra dimension,
so that a finite effective 4D Planck mass is obtained
even in non-compact space.
Since in the model with the conventional scalar kinetic term
in the RS-II setup,
the integral of the warp factor turned out to be divergent
(or unwanted naked singularities at some points
in the bulk arise)~\cite{kkl2,kkl3,nogo},
an exotic kinetic term $+1/(H_{MNPQ})^2$ was employed~\cite{kkl1}.
However, as shown in Refs.~\cite{dvali,dvalicc,newDGP},
if one restricts matter fields only on the brane, 
the 4D Newtonian potential could be obtained on the brane 
even with an infinite volume extra dimension
by introducing an additional brane-localized gravity kinetic term.  

The above observation of Ref.~\cite{dvali,newDGP} that 
the introduction of a brane-localized gravity term 
$\delta(y)M_4^2\sqrt{|\hat{g}_4|}\hat{R}_4$ 
can give a $-\frac1r$ potential (for a certain range of a parameter) 
triggered our reconsideration on the model with the conventional 
kinetic energy term of the three-form anti-symmetric tensor field 
$A_{MNP}$ under the RS-II setup.
While Ref.~\cite{dvalicc} tried to explain the vacuum energy
$(0.003\ {\rm eV})^4$ with a very small bulk gravity mass parameter,
in this paper we try to search for the possibility
that the self-tuning idea works in the model with the standard kinetic
term $-(H_{MNPQ})^2$ and the brane gravity term 
$\delta(y)M_4^2\sqrt{|\hat{g}_4|}\hat{R}_4$.
Especially, we are interested in a relatively large bulk gravity
mass parameter case.
In order to reproduce the low energy Einstein gravity in our setup, 
we consider a more general form of brane gravity term.  
The explicit verification of the self-tuning solution in this paper
is composed of several ingredients.
They will be summarized in the conclusion.

\section{Self-tuning of the cosmological constant}

We consider the following action in 5D space-time $(x^\mu,y)$, 
\begin{eqnarray}
S&=&\int d^4xdy\bigg[\sqrt{|g_5|}\bigg(\frac{M_5^3}{2}R_5
-\frac{1}{2}\partial_M\phi\partial^M\phi+{\cal L}_m^b-\Lambda_b\bigg)
\nonumber \\
&&~~~~~+\delta(y)\sqrt{|\bar{g}_4|}\bigg(\frac{M_4^2}{2}\bar{R}_4
+{\cal L}_m^1-\Lambda_1\bigg) \bigg]~,~~\label{action}
\end{eqnarray}
where $g_5\equiv{\rm Det}g_{MN}$, $M,N=0,1,2,3,5$, 
and $\bar{g}_4={\rm Det}\bar{g}_{\mu\nu}$, $\mu,\nu=0,1,2,3$.   
$\Lambda_{b}$
($\Lambda_1$) indicates the bulk (brane) cosmological constant.
$\bar{R}_4$ is the 4D Ricci scalar $\bar{g}^{\mu\nu}\bar{R}^\rho_{\mu\rho\nu}$ 
(4D Einstein-Hilbert term).  
$\phi$ is a 5D massless scalar field.
To protect the massless scalar from radiative large mass generation,
it is desirable to regard $\phi$
as the dual field of a 5D 3 form tensor field $A_{MNP}$. The ordinary
standard model (SM) matter fields are regarded as brane fields living
only in the brane, and ${\cal L}_m^1$ (${\cal L}_m^b$) denotes the Lagrangian
describing SM (bulk) matter dynamics. We introduce a $Z_2$ symmetry in
the extra space, and assign even (odd) parities to $g_{\mu\nu}$,
$g_{55}$ ($g_{\mu 5}$, $\phi$) under $y\leftrightarrow -y$.
The presence of brane terms proportional to $\delta(y)$
in Eq.~(\ref{action}) explicitly breaks the 5D general covariance
into the 4D one. The introduction of such a 4D Einstein-Hilbert term
does not spoil any given symmetry~\cite{dvali,dvalicc,newDGP,ks}.
We suppose that $M_4$ is {\it comparable} to $M_5$.

In general, the brane-localized Einstein-Hilbert term
$\delta(y)M_4^2\sqrt{|\bar{g}_4|}\bar{R}_4$ can be constructed 
with a metric tensor relatively different from 
the bulk metric by a scale factor~\cite{newDGP},  
\begin{eqnarray} \label{diffmetric}
\bar{g}_{\mu\nu}(x,y)=g_{\mu\nu}(x,y)\times\Omega^{-2}(x,y)~,  
\end{eqnarray}     
because it preserves the 4D general covariance on the brane.   
$\Omega(x,z)$ can not be removed from the action 
by redefining the metric.   
Actually, the brane metric $\bar{g}_{\mu\nu}$ 
could be further generalized to    
\begin{eqnarray} \label{alpha}  
\bar{g}_{\mu\nu}\longrightarrow
\hat{g}_{\mu\nu}\equiv\bar{g}_{\mu\nu}+\frac{\alpha}{M_4^2}
\partial_y^2\bar{g}_{\mu\nu} ~,    
\end{eqnarray} 
replacing $\delta(y)M_4^2\sqrt{|\bar{g}_4|}\bar{R}_4[\bar{g}_{\mu\nu}]$ 
in Eq.~(\ref{action}) by a more extended form,  
$\delta(y)M_4^2\sqrt{|\hat{g}_4|}\hat{R}_4[\hat{g}_{\mu\nu}]$.  
Here $\alpha$ is a dimensionless coupling.  
A 4D general coordinate transformation still can be defined as
$\hat{g}'_{\mu\nu}=\frac{\partial x^\rho}{\partial x^{'\mu}}
\frac{\partial x^\sigma}{\partial x^{'\nu}}\hat{g}_{\rho\sigma}$
with $\partial_y^2(\frac{\partial x^\rho}{\partial x^{'\mu}})|_{y=0}=0$, 
and the $Z_2$ symmetry is preserved.  
Such corrections need to be considered in order to more precisely reproduce 
the 4D low energy Einstein gravity on the brane.  
At linearized level, as will be shown in section 3,  
$\Omega (x,y)$ in Eq.~(\ref{diffmetric}) 
plays an essential role to obtain the desirable graviton tensor structure,  
and the $\alpha$ correction is also necessary 
for the Newtonian potential at {\it low} energies~\cite{newDGP}.   
But they do not affect the background 5D Einstein equation.  

We take the ansatze for the metric tensor and scalar fields as
\begin{eqnarray}
&&ds^2=\beta^2(y)~\eta_{\mu\nu}dx^\mu dx^\nu+dy^2 ~,
~~\Omega(y)=\beta(y) ~,  \label{az1}\\
&&~~~~~~~~~~~~\partial_y\phi= \frac{\sqrt{2A}}{\beta^4}~,
~~\partial_\mu\phi=0 ~, \label{az2}
\end{eqnarray}
where $\eta_{\mu\nu}={\rm diag.}(-1,1,1,1)$,
and $A$ is a constant.
Thus, $\bar{g}_{\mu\nu}$ and $\hat{g}_{\mu\nu}$    
in Eqs.~(\ref{diffmetric}) and (\ref{alpha}) are just $\eta_{\mu\nu}$.  
Eq.~(\ref{az2}) trivially satisfies the scalar field equation
with Eq.~(\ref{az1}),
\begin{eqnarray}
\partial_M\bigg(\sqrt{-g_5}~\partial^M\phi\bigg)=0 ~.
\end{eqnarray}
With the metric ansatz (\ref{az1}), the 5D Einstein equation leads 
to~\cite{kkl2} 
\begin{eqnarray} \label{backeq1}
3\bigg(\frac{\ddot{\beta}}{\beta}\bigg)
+3\bigg(\frac{\dot{\beta}}{\beta}\bigg)^2
&=&-\frac{\Lambda_b}{M_5^3}-\delta(y)\frac{\Lambda_1}{M_5^3}
-\frac{A}{M_5^3}~\frac{1}{\beta^8} ~, \\
6\bigg(\frac{\dot{\beta}}{\beta}\bigg)^2
&=&-\frac{\Lambda_b}{M_5^3}+\frac{A}{M_5^3}~\frac{1}{\beta^8} ~, 
\label{backeq2}
\end{eqnarray}
where the dots denote differentiations with respect to $y$, and
\begin{eqnarray} \label{para}
\lambda_b\equiv\sqrt{\frac{|\Lambda_b|}{6M_5^3}}~~,
~~~\lambda_1\equiv\frac{\Lambda_1}{6M_5^3}~~, ~~~{\rm
and}~~~a\equiv\sqrt{\frac{A}{6M_5^3}} ~.
\end{eqnarray}
Note that the brane-localized Einstein-Hilbert term does not contribute to
the 5D Einstein equation Eq.~(\ref{backeq1}) 
with the metric ansatz Eq.~(\ref{az1}).

Depending on the sign of $\Lambda_b$, we have three classes of
bulk solutions~\cite{kkl2}
\begin{eqnarray} 
&&\Lambda_b>0 : \ \
\beta(y)=\bigg(\frac{a}{\lambda_b}\bigg)^{1/4}\bigg[ {\rm sin}
(4\lambda_b|y|+c_+)\bigg]^{1/4}
~~~~~~~ \label{sol1}\\
&&\Lambda_b<0 :\ \
\beta(y)=\bigg(\frac{a}{\lambda_b}\bigg)^{1/4}\bigg[{\rm sinh}
(4\lambda_b|y|+c_-)\bigg]^{1/4} ~~~~~\label{sol2}\\
&&\Lambda_b=0 :\ \ \beta(y)=\bigg[4a|y|+c\bigg]^{1/4}
~~~~~~~~~~~~~~~~~ \label{sol3}
\end{eqnarray}
where the integration constants $c_+$, $c_-$. and $c$ should be
determined by the boundary conditions:
$[\beta^\prime/\beta]_{0^+}=-\lambda_1$.
To avoid naked
singularities at some points in the bulk, we should take only the
solutions for $\Lambda_b<0$ and $\Lambda_b=0$ with {\it positive}
values of $c$ and $c_-$,
\begin{eqnarray} \label{self1}
c_-=-{\rm coth}^{-1}\bigg(\frac{\lambda_1}{\lambda_b}\bigg)~>0 ~~~
&&{\rm for}~~\Lambda_b<0 ~,
\\  \label{self2}
c=-\bigg(\frac{a}{\lambda_1}\bigg)~>0 ~~~~~~~~~~~~&&{\rm
for}~~\Lambda_b=0 ~,
\end{eqnarray}
Hence, the brane vacuum energy is required to be negative for
$\Lambda_b=0$ and $\Lambda_b<0$. In the boundary conditions
Eqs.~(\ref{self1}) and (\ref{self2}), we note that $\lambda_1$ is
never fixed by any Lagrangian parameter. Any arbitrary negative
value of $\lambda_1$ allows a 4D flat space-time solution, by
adjusting $c_-$ in Eq.~(\ref{self1}), and $c$ and $a$ in
Eq.~(\ref{self2}) such that the boundary conditions at the brane
are fulfilled. Even if the electroweak and QCD phase transitions 
change the brane vacuum energy $\lambda_1\rightarrow\lambda_1'$, 
the flat 4D space-time could be maintained by proper shifts
$c_{(-)}\rightarrow c_{(-)}'$ and $a\rightarrow a'$. It means that
the profiles of the metric tensor and the scalar fields in the extra
dimension are changed. Since they are dynamical fields, it is
always possible.
We note that our self-tuning mechanism is not associated
with any parameter sensitive to low energy physics~\cite{nilles2}.
As will be turn out, the 4D Newtonian constant
is given by $M_4$ rather than $M_5$ in our model.

According to Refs.~\cite{hawking}, 
the flat universe is quantum mechanically most (infinitely) probable 
in 4D space-time, even though a 4D dS or AdS universe 
is also classically allowed [In fact, the 4D dS and AdS solutions exist 
in our model].     
Hence, the flat universe can always be {\it dynamically} chosen 
through the self-tuning mechanism~\cite{kkl2,selfinf}.  
%
%

From Eq.~(\ref{az2}), and with the solutions Eqs.~(\ref{sol2}) 
and (\ref{sol3}), one can easily find the background solution of $\phi_0(y)$, 
\begin{eqnarray}
%
&&\Lambda_b<0 :\ \  \label{sol<0}
\phi_0(y)=\frac{\sqrt{12M_5^3}}{4}~{\rm ln}\bigg[
\frac{{\rm tanh}\bigg(\frac{4\lambda_b|y|+c_-}{2}\bigg)}
{{\rm tanh}\frac{c_-}{2}}\bigg]\times{\rm sgn}(y)
\label{phi2} \\  \label{sol=0}
&&\Lambda_b=0 :\ \ \phi_0(y)=\frac{\sqrt{12M_5^3}}{4}~{\rm ln}\bigg[
\frac{4a}{c}|y|+1\bigg]\times{\rm sgn}(y)\label{phi3} ~,   
\end{eqnarray}
where ${\rm sgn}(y)$ is defined as $+1$ ($-1$) for $y>0$ ($y<0$). 
Note that $\phi$ has the odd parity under $y\leftrightarrow -y$.  
Since it is a continuous function at $y=0$, 
$\partial_y\phi_0$ does not induce the delta function at $y=0$, which is 
consistent with Eq.~(\ref{az2}).   

With the 5D gravity action, the gravity potential $-\frac{1}{r^2}$
is naively expected in non-compact 5D space-time.
In Refs.~\cite{kkl2,kkl3,cho,nilles},
two branes are employed to make the extra dimension compact.
However, the introduction of two branes turns out to accompany
a fine-tuning relation between two brane cosmological constants
if the interval length between two branes must be somehow 
stabilized~\cite{nilles2}.
In this paper, instead of employing two branes,
we introduce just one brane to avoid the fine-tuning, and
an additional brane-localized Einstein-Hilbert term to get a realistic
$-\frac{1}{r}$ gravity potential.
Although there is no normalizable zero mode of the 5D graviton
in general in this setup~\cite{nogo},  
the introduction of a brane-localized Einstein-Hilbert term provides
a desirable 4D gravity interaction on the brane embedded
in non-compact extra dimension~\cite{dvali,newDGP,shifman}.
To see it explicitly, we need to consider metric perturbations
near the background solutions Eqs.~(\ref{sol3}) and (\ref{sol2}), and
examine how the graviton propagator is given in this setup.

\section{Metric perturbation}

The relevant perturbed metric near the above background solutions is
\begin{eqnarray}
&&ds^2=\bigg[\beta^2+\delta\beta^2(x,y)\bigg]\bigg(\eta_{\mu\nu}
+\delta g_{\mu\nu}(x,y)\bigg)dx^\mu dx^\nu
+\bigg(1+\delta g_{55}(x,y)\bigg)dy^2
\\
&&\equiv\bigg[\frac{1}{K^2(z)}\bigg(1+F(x,z)\bigg)\bigg]
\bigg[\bigg(\eta_{\mu\nu}+h_{\mu\nu}(x,z)\bigg)dx^\mu dx^\nu
+\bigg(1+h_{55}(x,z)\bigg)dz^2\bigg]~,~~~~~~~\label{ds2} 
\nonumber 
\end{eqnarray}
where $|z|=\int^z_0 dy/\beta(y)$, $K(z)=1/\beta(y(z))$.
For convenience of the analysis, 
we change the $(x,y)$ coordinate into $(x,z)$.  
$F$ ($<<1$) denotes perturbation of the warp factor,  
while $h_{\mu\nu}$ is interpreted as the 4D graviton.  
Small energy-momentum tensors by brane and bulk matter
cause such small metric perturbations.    
Since the brane metric $\bar{g}_{\mu\nu}$ can be relatively different 
from the bulk metric $g_{\mu\nu}$ by $\Omega^2$ 
($\approx [1+\omega(x,z)]/K^2$),   
we can always redefine $h_{\mu\nu}$ 
[or $\eta^{\mu\nu}h_{\mu\nu}$ ($\equiv h$)]  
and $F$ such that on the brane 
$\bar{g}_{\mu\nu}\approx \eta_{\mu\nu}+\bar{h}_{\mu\nu}$,  
while in the bulk $g_{\mu\nu}\approx (\eta_{\mu\nu}+\bar{F}\eta_{\mu\nu}
+\bar{h}_{\mu\nu})/K^2$ 
with $\bar{h}_{\mu\nu}=h_{\mu\nu}+(F-\omega)\eta_{\mu\nu}$ 
(or $\bar{h}=h+4(F-\omega)$) and $\bar{F}=\omega$~\cite{newDGP}.  
From now on we will drop the ``bar''s.  
%
%
We define $G$ such that $g_{55}\approx (1+G)/K^2$ (i.e. $G=F+h_{55}$).  
We will consider the $\alpha$ correction of Eq.~(\ref{alpha}) later,  
because it just affects the boundary condition at $z=0$.  

For a perturbed metric $g_{MN}\approx(\eta_{MN}+H_{MN})/K^2(z)$,
the linearized 5D Einstein tensor is generally given by~\cite{kkl3}
\begin{eqnarray}
&&\delta G^b_{MN}=-\frac{1}{2}\nabla_5^2\overline{H}_{MN}
+\partial_{(M}\partial^P\overline{H}_{N)P}
-\frac{1}{2}\eta_{MN}\partial^P\partial^Q\overline{H}_{PQ}
\nonumber \\
&&~~~~~~~-\frac{3K'}{2K}\bigg[\partial_MH_{N5}+\partial_NH_{M5}
-\partial_zH_{MN}\bigg] 
\\
&&+3\eta_{MN}\bigg[\left(\frac{K''}{K}
-2\frac{K'^2}{K^2}\right)H_{55}
+\frac{K'}{K}\partial^P\overline{H}_{P5}\bigg]
-3\bigg[\frac{K''}{K}-2\frac{K'^2}{K^2}\bigg]H_{MN}~,
\nonumber
\end{eqnarray}
where $(M,N)$ is a half of the symmetric combination, 
$\nabla_5^2$ indicates
$\eta^{\mu\nu}\partial_{\mu}\partial_{\nu} +\partial_z^2$ and 
$'$ denotes the derivative with respect to $z$.  
$\overline{H}_{MN}$ is defined as 
$H_{MN}-\frac{1}{2}\eta_{MN}(\eta^{\mu\nu}H_{\mu\nu}+H_{55})$.
In our case, $H_{\mu\nu}$ and $H_{55}$ are given by     
\begin{eqnarray}
H_{\mu\nu}=F\eta_{\mu\nu}+h_{\mu\nu}~,~~~{\rm and}~~~H_{55}=G~.
\end{eqnarray}
On the other hand, the contribution by the localized gravity term 
in Eq.~(\ref{action}) to the linearized Einstein tensor is 
\begin{eqnarray} \label{G^1}
\delta G^1_{\mu\nu}=-\frac{M_4^2}{M_5^3}\delta(z)K\bigg[
\frac{1}{2}\nabla_4^2\bar{h}_{\mu\nu}
-\partial_{(\mu}\partial^\lambda\bar{h}_{\nu)\lambda}
+\frac{1}{2}\eta_{\mu\nu}\partial^\lambda\partial^\delta
\bar{h}_{\lambda\delta}\bigg] ~, 
\end{eqnarray}
where $\nabla_4^2\equiv\eta^{\mu\nu}\partial_{\mu}\partial_{\nu}$, and 
$\bar{h}_{\mu\nu}\equiv h_{\mu\nu}-\frac{1}{2}\eta_{\mu\nu}h$. 
The (linearized) energy-momentum tensor turns out to be~\cite{kkl3}  
\begin{eqnarray} \label{tmn}
&&\delta T_{\mu\nu}=-3\bigg[\frac{K''}{K}-2\frac{K'^2}{K^2}\bigg]H_{\mu\nu} 
+\frac{\eta_{\mu\nu}}{2M_5^3}\bigg[(\phi_0')^2
+\frac{\Lambda_1}{K}\delta(z)\bigg]G
\\
&&~~~~~~~~~-\frac{\eta_{\mu\nu}}{M_5^3}\phi_0'\varphi'
+\frac{1}{M_5^3}T^b_{\mu\nu}+\frac{1}{M_5^3}\delta(z)KT^1_{\mu\nu} ~, 
\nonumber \\
&&\delta T_{\mu 5}=-3\bigg[\frac{K''}{K}-2\frac{K'^2}{K^2}\bigg]h_{\mu 5}
+\frac{1}{M_5^3}\phi_0'\partial_\mu\varphi+\frac{1}{M_5^3}T^b_{\mu5} ~,
\label{tm5} \\
&&\delta T_{55}=\bigg[6\frac{K^{'2}}{K^2}-\frac{1}{2}(\phi_0')^2\bigg]G
+\frac{1}{M_5^3}\phi_0'\varphi'+\frac{1}{M_5^3}T^b_{55}~,~
\label{t55}
\end{eqnarray}
where we utilized $\delta(z)h_{\mu 5}=0$ and the background Einstein equations
Eqs.~(\ref{backeq1}) and (\ref{backeq2}), which are
translated in the $(x^\mu,z)$ coordinate into~\footnote{
In Eq.~(\ref{tm5}) we have corrected an error found in the corresponding 
expression of Ref.~\cite{kkl3}.  
}  
\begin{eqnarray} \label{backeq1'}
K^2\bigg[\frac{K''}{K}-2\frac{K^{'2}}{K^2}\bigg]&=&\frac{1}{3M_5^3}\bigg[
\Lambda_b+\delta(z)K\Lambda_1+AK^8\bigg] ~,  \\
K^2\bigg(\frac{K'}{K}\bigg)^2&=&-\frac{1}{6M_5^3}\bigg[\Lambda_b-AK^8\bigg] ~.
\label{backeq2'}
\end{eqnarray}
In Eqs.~(\ref{tmn})--(\ref{t55}), $\varphi(x,z)$ denotes 
the linearly perturbed bulk scalar
near the background solution $\phi_0(z)$ given
in Eq.~(\ref{sol<0}) or (\ref{sol=0}).
Thus, the odd parity should be assigned to it.
$T^b_{MN}$ and $T^1_{\mu\nu}$ in Eqs.~(\ref{tmn})--(\ref{t55})
denote the energy-momentum tensors by bulk and
brane matter.  For simplicity we will neglect $T^b_{MN}$ in this paper.
The standard model matter fields are assumed to contribute to 
$T^1_{\mu\nu}$.
Note that the resultant linearized Einstein equation 
$\delta G^b_{MN}+\delta G^1_{\mu\nu}\delta^\mu_M\delta^\nu_N=\delta T_{MN}$ 
is invariant under the 4D gauge transformation,
$h_{\mu\nu}'=h_{\mu\nu}+\partial_\mu\xi_\nu+\partial_\nu\xi_\mu$
and $h_{\mu 5}'=h_{\mu 5}+\partial_z\xi_{\mu}$. 
Thus, we can freely choose a proper gauge condition for $h_{\mu\nu}$. 
With the gauge $\partial^\mu\bar{h}_{\mu\nu}=0$, 
which fixes the gauge parameter $\xi_{\mu}(x,z)$, 
the linearized Einstein equation reads 
\newpage 
\begin{eqnarray}
(\mu\nu)&:&  \label{mn}
~-\bigg(\partial_\mu\partial_\nu-\eta_{\mu\nu}\nabla_4^2\bigg)\bigg[
F+\frac{1}{2}G\bigg]+\frac{3}{2}\eta_{\mu\nu}\partial_z^2F 
\nonumber \\
&&~
+3\eta_{\mu\nu}\bigg[\bigg(\frac{K''}{K}-2\frac{K^{'2}}{K^2}\bigg)G
-\frac{3K'}{2K}\partial_zF+\frac{K'}{2K}\partial_zG\bigg]
\nonumber \\
&&~-\frac{1}{2}\bigg(\nabla_5^2-3\frac{K'}{K}\partial_z\bigg)
\bigg[h_{\mu\nu}-\frac{1}{2}\eta_{\mu\nu}h\bigg]
+\frac{1}{4}\eta_{\mu\nu}\bigg[\partial_z^2h-3\frac{K'}{K}\partial_zh\bigg]
%
%
 \\
&&~+\bigg(\partial_z-3\frac{K'}{K}\bigg)\bigg[
\partial_{(\mu}h_{\nu)5}-\eta_{\mu\nu}\partial^\lambda h_{\lambda 5}\bigg]
-\frac{M_4^2}{2M_5^3}\delta(z)K\nabla_4^2\bigg[
h_{\mu\nu}-\frac{1}{2}\eta_{\mu\nu}h\bigg]
\nonumber \\
&&
=\frac{\eta_{\mu\nu}}{2M_5^3}\bigg[(\phi_0')^2
+\frac{\Lambda_1}{K}\delta(z)\bigg]G
-\frac{\eta_{\mu\nu}}{M_5^3}\phi_0'\varphi'
+\frac{1}{M_5^3}\delta(z)KT^1_{\mu\nu}
%
~,
~~~~~\nonumber \\
(\mu 5)&:&  \label{m5}
-\frac{3}{2}\partial_\mu\partial_z\bigg[F+\frac{1}{6}h\bigg]
-\frac{3K'}{2K}\partial_\mu G -\frac{1}{2}\bigg[
\nabla_4^2h_{\mu 5}-\partial_\mu\partial^\lambda h_{\lambda 5}\bigg]
=\frac{1}{M_5^3}\phi_0'\partial_\mu\varphi
~, 
\\
(55)&:&   \label{55}
\frac{3}{2}\nabla_4^2\bigg[F+\frac{1}{6}h\bigg]-6\frac{K'}{K}\partial_z\bigg[
F+\frac{1}{4}h\bigg]+3\frac{K'}{K}\partial^\lambda h_{\lambda 5} 
=-\frac{\Lambda_b}{M_5^3}\frac{G}{K^2}
+\frac{1}{M_5^3}\phi_0'\varphi'
~,~~~~~~ 
\end{eqnarray}
where the singular terms proportional to $\delta(z)K(z)$ ($=\delta(y)$)  
in the left hand side of Eq.~(\ref{mn}) came from Eq.~(\ref{G^1}).  
%

In principle, the case with $\Lambda_b<0$ as well as with $\Lambda_b=0$ is
workable. 
But from now on, we will focus our discussion only on the $\Lambda_b=0$ case, 
because this case gives an exact solution.  
We found that the solutions of $\varphi$, $G$, $F$ are  
\begin{eqnarray} \label{solphi}
&&\varphi(x,z)=\sqrt{\frac{M_5^3}{48}}~\bigg[h(x,z)-h^0(x)\bigg]
\times{\rm sgn}(z) ~,  \\  \label{sol55}
&&G(x,z)=\frac{1}{3}\bigg[h(x,z)-h^0(x)\bigg] ~, \\
&&F(x,z)=-\frac{1}{6}h(x,z)~,  \label{solF}
\end{eqnarray}
where $h^0(x)\equiv h(x,z=0)$.  
%
%
%
%
The bulk solution of $h_{\mu 5}$ is given by 
$h_{\mu 5}(x,z)=\partial_\mu\Psi(x)K^3(z)$, 
where $\Psi(x)$ fulfills $\nabla_4^2\Psi(x)=0$. 
Hence, $\partial^\mu h_{\mu 5}=\nabla_4^2h_{\mu 5}=0$.     
Even if there exists the bulk solution, however, 
it can not satisfy the boundary condition, $h_{\mu 5}|_{z=0}=0$  
unless $\Psi(x)=0$.   
One can check that Eqs.~(\ref{solphi}), (\ref{sol55}), (\ref{solF}) fulfill 
Eqs.~(\ref{mn})--(\ref{55}) using the relations  
\begin{eqnarray} \label{rel1}
&&~~~~~\phi_0'=\sqrt{2A}~K^3 ~, 
\\ \label{rel2}
&&\frac{K'}{K}{\rm sgn}(z)=-\sqrt{\frac{A}{6M_5^3}}~K^3 ~,
\end{eqnarray}
which came from Eqs.~(\ref{az2}) and (\ref{backeq2'}), respectively.  
Note that $\varphi$ in Eq.~(\ref{solphi}) is continuous at $z=0$, and so 
$\partial_z\varphi\propto{\rm sgn}(z)\partial_zh$.    

By Eqs.~(\ref{solphi}), (\ref{sol55}), (\ref{solF}), 
and Eqs.~(\ref{backeq1'}), (\ref{rel1}), (\ref{rel2}),  
the $(\mu\nu)$ component Eq.~(\ref{mn}) becomes much simpler,   
\begin{eqnarray} 
&&-\frac{1}{2}\bigg(\nabla_5^2-3\frac{K'}{K}\partial_z\bigg)\bigg[
h_{\mu\nu}-\frac{1}{2}\eta_{\mu\nu}h\bigg]
-\frac{M_4^2}{2M_5^3}\delta(z)K\nabla_4^2\bigg[h_{\mu\nu}
-\frac{1}{2}\eta_{\mu\nu}h\bigg] 
\nonumber \\ 
&&~~~~~
~~~~~~~~~~=-\frac{1}{6}\bigg(
\partial_\mu\partial_\nu-\eta_{\mu\nu}\nabla_4^2\bigg)h^0
%
+\frac{1}{M_5^3}\delta(z)KT^1_{\mu\nu} ~, 
\label{simplemn}
\end{eqnarray}
where $\delta(z)G(x,z)=0$ was used.    
Eq.(\ref{simplemn}) identically satisfies 
the energy-momentum conservation law $\partial^\mu T^1_{\mu\nu}=0$.\footnote{
Only with Eqs.~(\ref{mn}), (\ref{m5}), and (\ref{rel1}), one can show
that $\partial^\mu T^1_{\mu\nu}=0$ is generally satisfied.
} 
The trace of Eq.~(\ref{simplemn}) gives 
\begin{eqnarray}\label{traceeq}
\bigg(\nabla_5^2-3\frac{K'}{K}\partial_z\bigg)\bigg[h(x,z)-h^0(x)\bigg]
+\frac{M_4^2}{M_5^3}\delta(z)K\nabla_4^2h^0(x) 
=\frac{2}{M_5^3}\delta(z)KT^1(x) ~, 
\end{eqnarray}
where $T^1\equiv \eta^{\mu\nu}T^1_{\mu\nu}$.   
%
In Eq.~(\ref{traceeq}) we note that in the bulk, $h(x,z)-h^0(x)$ 
($\equiv H(x,z)$) rather than $h(x,z)$ seems a dynamical field. 
At $z=0$, the equation of motion for the trace mode is 
$\nabla_4^2h^0(x)=\frac{2}{M_4^2}T^1(x)$, 
whereas at $z\neq 0$, dynamics of $H(x,z)$ is  
described with $[\nabla_5^2-3\frac{K'}{K}\partial_z]H(x,z)=0$.\footnote{
Similarly the traceless part of Eq.~(\ref{simplemn}) can be recast into
\begin{eqnarray}
\bigg(\nabla_5^2-3\frac{K'}{K}\partial_z\bigg)\bigg[
h^T_{\mu\nu}(x,z)-\bigg(\partial_\mu\partial_\nu
-\frac{1}{4}\eta_{\mu\nu}\nabla_4^2\bigg)f(x)\bigg]
+\frac{M_4^2}{M_5^3}\delta(z)\nabla_4^2h^0_{\mu\nu}(x,z)
=-\frac{2}{M_5^3}\delta(z)KT^{1T}_{\mu\nu}(x) ~,
\nonumber
\end{eqnarray}
where the superscript $T$ denotes the traceless part of the corresponding 
tensor field, and $\nabla_4^2f\equiv \frac{1}{3}h^0$.  
}
Using Eq.~(\ref{traceeq}) and 
${\rm sgn}(z)\delta(z)~(=\frac{1}{4}\partial_z[({\rm sgn}(z))^2])~=0$ 
together with Eqs.~(\ref{solphi}), (\ref{sol55}) and (\ref{solF}), 
one can easily check that the linearized scalar equation 
for $\varphi$~\cite{kkl3}, 
\begin{eqnarray} \label{linearscalar}
\bigg(\nabla_5^2-3\frac{K'}{K}\partial_z\bigg)\varphi
+\frac{\phi_0'}{2}\bigg(4F'+h'-G'\bigg)=0
\end{eqnarray}
is satisfied.\footnote{
The consistency between the trace of Eqs.~(\ref{mn}) and (\ref{linearscalar}) 
compelled us to choose the ``gauge'' $\partial^\mu h_{\mu 5}=0$, 
which is reminiscent of the gauge choice in massive $U(1)$ gauge theory.  
It is because the background solution of $g_{MN}$ breaks 
the 5D Lorentz symmetry into the 4D one.  
}  
After some algebra, Eq.~(\ref{simplemn}) takes the following form, 
\begin{eqnarray} \label{maineq}
&&~~~~~~~~\bigg(\nabla_5^2-3\frac{K'}{K}\partial_z\bigg)h_{\mu\nu}
+\frac{M_4^2}{M_5^3}\delta(z)K\nabla_4^2h_{\mu\nu}
\\
&&=-\frac{2}{M_5^3}\bigg(S_{\mu\nu}
-\frac{1}{2}\eta_{\mu\nu}S\bigg)
-\frac{2}{M_5^3}\delta(z)K\bigg(
T^1_{\mu\nu}-\frac{1}{2}\eta_{\mu\nu}T^1\bigg) ~, 
\nonumber 
\end{eqnarray}
where $S_{\mu\nu}\equiv -\frac{M_5^3}{6}(\partial_\mu\partial_\nu
-\eta_{\mu\nu}\nabla_4^2)h^0$, and 
$S\equiv \eta^{\mu\nu}S_{\mu\nu}$.    

Now let us solve Eq.~(\ref{maineq}).
For $\Lambda_b=0$, $z$ and $K(z)$ are calculated to give
\begin{eqnarray}
&&|z|=\int^y_0\frac{dy}{(4a|y|+c)^{1/4}}
=\frac{1}{3a}\bigg[(4a|y|+c)^{3/4}-c^{3/4}\bigg] ~, \\
&&K(z)=\frac{1}{\beta(y(z))}=(3a|z|+c^{3/4})^{-1/3} ~.   
\label{K}
\end{eqnarray}
Eq.~(\ref{K}) is consistent with Eq.~(\ref{rel2}).    
Note that the brane position $y=0$ corresponds to $z=0$. 
In 4D momentum space $(p,z)$, Eq.~(\ref{maineq}) becomes
\begin{eqnarray} \label{eom}
&&~~~~~~~~\bigg[k^2+\partial_z^2
+\frac{3a{\rm sgn}(z)}{(3a|z|+c^{3/4})}\partial_z
+\frac{M_4^2}{M_5^3}\delta(z)c^{-1/4}k^2\bigg]\tilde{h}_{\mu\nu}(k,z)
\\
&&=-\frac{2}{M_5^3}\bigg(\tilde{S}_{\mu\nu}(k)
-\frac{1}{2}\eta_{\mu\nu}\tilde{S}(k)\bigg)
-\frac{2c^{-1/4}}{M_5^3}\delta(z)\bigg(\tilde{T}^{1}_{\mu\nu}(k)
-\frac{1}{2}\eta_{\mu\nu}\tilde{T}^1_{\mu\nu}(k)\bigg) ~,
\nonumber 
\end{eqnarray}
where $k\equiv\sqrt{-p^\mu p_\mu}$ ($\ge 0$), and the fields with tildes 
indicate the 4D Fourier-transformed fields.     
%
%
The solution of Eq.~(\ref{eom}) is given by Bessel functions of order zero. 
The traceless and trace parts of the bulk solution are given by  
\begin{eqnarray} \label{bulksolless}
\tilde{h}^T_{\mu\nu}(k,z)=-\frac{2\tilde{S}^T_{\mu\nu}(k)}{M_5^3k^2}
+p_{\mu\nu}(k)
\bigg(J_0[k(|z|+c^{3/4}/3a)]+u(k)Y_0[k(|z|+c^{3/4}/3a)]\bigg),~
\\ \label{bulksoltrace}
\tilde{H}(k,z)\equiv \tilde{h}(k,z)-\tilde{h}^0(k)=
q(k)\bigg(J_0[k(|z|+c^{3/4}/3a)]+v(k)Y_0[k(|z|+c^{3/4}/3a)]\bigg).~   
\end{eqnarray}
In Eq.~(\ref{bulksolless}) 
$\tilde{h}^T_{\mu\nu}$, $\tilde{S}_{\mu\nu}^T$ denote 
the traceless parts of $\tilde{h}_{\mu\nu}$, $\tilde{S}_{\mu\nu}$, i.e.    
$\tilde{h}^T_{\mu\nu}\equiv \tilde{h}_{\mu\nu}
-\frac{1}{4}\eta_{\mu\nu}\tilde{h}$ and  
$\tilde{S}^T_{\mu\nu}\equiv\tilde{S}_{\mu\nu}
-\frac{1}{4}\eta_{\mu\nu}\tilde{S}$.  
Similarly we define $\tilde{T}^{1T}_{\mu\nu}\equiv\tilde{T}^1_{\mu\nu}
-\frac{1}{4}\eta_{\mu\nu}\tilde{T}$.  
%
%
$p_{\mu\nu}(k)$, $q(k)$ and $u(k)$, $v(k)$ should be determined 
by boundary conditions.  

Eq.~(\ref{G^1}) could be supplemented by the higher derivative correctional 
terms,  
\begin{eqnarray} \label{new}
\delta G^{1}_{\mu\nu}\supset-\frac{\alpha}{M_5^3}\delta(z)K\partial_z^2\bigg[
\frac{1}{2}\nabla_4^2\bar{h}_{\mu\nu}
-\partial_{(\mu}\partial^\lambda\bar{h}_{\nu)\lambda}
+\frac{1}{2}\eta_{\mu\nu}\partial^\lambda\partial^\delta
\bar{h}_{\lambda\delta}\bigg] ~, 
\end{eqnarray}
which was seriously considered in Ref.~\cite{newDGP}.     
Here $\alpha$ is a dimensionless coupling.  
Note that Eq.~(\ref{new}) is consistent 
with all the symmetries considered so far.  
Invariance of Eq.~(\ref{new}) under 
$h'_{\mu\nu}=h_{\mu\nu}+\partial_\mu\xi_\nu+\partial_\nu\xi_\mu$  
enables us to impose the gauge condition chosen above. 
%
Thus, it reduces to $\frac{-\alpha}{2M_5^3}\delta(z)K\nabla_4^2\partial_z^2[
h_{\mu\nu}-\frac{1}{2}\eta_{\mu\nu}h]$, and so $M_4^2$ in Eqs.~(\ref{mn}), 
(\ref{simplemn}), (\ref{traceeq}), (\ref{maineq}), and (\ref{eom}) 
should be replaced by $M_4^2(1+\frac{\alpha}{M_4^2}\partial_z^2)$.   

Eq.~(\ref{new}) can be effectively obtained from Eq.~(\ref{G^1}) by redefining 
$h_{\mu\nu}\rightarrow 
\hat{h}_{\mu\nu}\equiv h_{\mu\nu}+\frac{\alpha}{M_4^2}\partial_z^2h_{\mu\nu}$ 
only on the brane.    
Hence, Eq.~(\ref{new}) could appear in the linearized equation 
derived from the Lagrangian 
with a more extended brane-localized term     
$\delta(z)KM_4^2\sqrt{|\hat{g}_4|}\hat{R}_4[\hat{g}_{\mu\nu}]$,  
which is constructed with a ``brane metric'',   
$\hat{g}_{\mu\nu}\equiv\bar{g}_{\mu\nu}+\frac{\alpha}{M_4^2}
\partial_z^2\bar{g}_{\mu\nu}$~\cite{newDGP}.    
%
%
In that case, the brane-localized energy-momentum tensor should be 
coupled to $\hat{g}_{\mu\nu}$, which leads to a gravity interaction term 
in the linearized Lagrangian, 
$\delta(z)\frac{1}{2}\hat{h}^{\mu\nu}T^1_{\mu\nu}$.  
Thus, the brane-localized gravity kinetic and interaction terms 
in the linearized Lagrangian is given by  
\begin{eqnarray}
{\cal L}^1_{\rm lin}=-\delta(z)K\bigg[
\frac{M_4^2}{4}\bigg(
\frac{1}{2}(\partial_\mu \hat{h}_{\nu\rho})^2
-\frac{1}{2}(\partial_\mu \hat{h})^2-(\partial^\nu \hat{h}_{\mu\nu})^2
+\partial_\mu \hat{h}\partial_\nu \hat{h}^{\mu\nu}
\bigg)-\frac{1}{2}\hat{h}^{\mu\nu}T^1_{\mu\nu}\bigg] ~.
\end{eqnarray}   
However, the bulk gravity kinetic term is still constructed with $g_{MN}$ 
as discussed earlier.   

Indeed, due to the presence of the $\alpha$ terms, 
the variation of the linearized Lagrangian  
with respect to $h_{\mu\nu}$   
turns out to give a constraint equation~\cite{newDGP}, 
\begin{eqnarray} \label{constraint}
\delta G^1_{\mu\nu}=\frac{1}{M_5^3}\delta(z)KT^1_{\mu\nu} ~,    
\end{eqnarray} 
as well as the usual linearized Einstein equation,  
$\delta G^b_{\mu\nu}+\delta G^1_{\mu\nu}
=\frac{1}{M_5^3}T^b_{\mu\nu}+\frac{1}{M_5^3}\delta(z)KT^1_{\mu\nu}$.  
It implies $\delta G^b_{\mu\nu}=\frac{1}{M_5^3}T^b_{\mu\nu}$ ($=0$). 
On the other hand, the variation with respect to $F$ turns out to give 
$\delta G^{b~M}_M=\frac{1}{M_5^3}T^{b~M}_M$ ($=0$), 
which is consistent with $\delta G^b_{\mu\nu}=\frac{1}{M_5^3}T^b_{\mu\nu}$ 
($=0$).  It is because $F$ contributes only to the bulk.  

Eq.~(\ref{new}) provides a new non-trivial boundary condition at $z=0$ 
with leaving intact the forms of the bulk solutions Eqs.~(\ref{bulksolless}) 
and (\ref{bulksoltrace});    
only smooth graviton wave functions near the brane are allowed, 
\begin{eqnarray} \label{bdycondi}
\partial_z\tilde{h}_{\mu\nu}|_{z=0^+}
=\partial_z\tilde{h}_{\mu\nu}|_{z=0^-}=0~.
\end{eqnarray}   
Otherwise a highly singular term proportional to $\delta^2(z)$ arises from 
$-\frac{\alpha}{2M_5^3}\delta(z)k^2\partial_z^2\tilde{h}_{\mu\nu}$, 
which can not be matched to the right hand side of Eq.~(\ref{eom}).     
As will be shown, the boundary condition Eq.~(\ref{bdycondi}) makes   
the effect by $\sqrt{|g_5|}M_5^3R_5$ suppressed on the brane 
without assuming small $M_5$~\cite{newDGP}.    
The $-\frac{1}{r}$ gravity potential on the brane is ensured 
by $\delta(z)KM_4^2\sqrt{|\hat{g}_4|}\hat{R}_4$.   
By Eq.~(\ref{bdycondi}), $u(k)$ and $v(k)$ in Eqs.~(\ref{bulksolless}) and 
(\ref{bulksoltrace}) are determined to be   
\begin{eqnarray} \label{uv}
u(k)=v(k)=-\frac{J_1[kc^{3/4}/3a]}{Y_1[kc^{3/4}/3a]}  ~.    
\end{eqnarray}
Since $\partial^2_z\tilde{h}^T_{\mu\nu}$, $\partial^2_z\tilde{h}$ do not 
induce any singular term proportional to $\delta(z)$ with Eq.~(\ref{uv}),   
we can fulfill $\delta G^b_{\mu\nu}=0$ (and so Eq.~(\ref{constraint})).   
The boundary condition at $z=0$ can be satisfied   
simply by setting equal the coefficients of the delta functions  
appearing in Eqs.~(\ref{eom}) and (\ref{new}).   

In fact, there does not exist the solution of $\tilde{H}(k,z)$ 
($\equiv\tilde{h}(k,z)-\tilde{h}^0(k)$) at $z\neq 0$,  
because only the vanishing $q(k)$ is consistent 
with the boundary conditions.    
However, $h^0(x)$ still can propagate 
in the longitudinal direction of the brane.     
At $z=0$, as discussed in Eq.~(\ref{traceeq}),         
\begin{eqnarray} \label{soltrace}
\tilde{h}^0(k)=\frac{2\tilde{T}^1(k)}{M_4^2k^2} ~. 
\end{eqnarray}
%
%
On the other hand, from Eqs.~(\ref{eom}) and (\ref{new}), 
the solution of $\tilde{h}^T_{\mu\nu}$ at $z=0$ is given by 
\begin{eqnarray} \label{solless}
\tilde{h}^T_{\mu\nu}(k,z=0)=-\frac{2}{M_4^2k^2}
\bigg[\tilde{T}^{1T}_{\mu\nu}(k)+O\bigg(
\frac{\alpha k^2\tilde{T}^{1T}_{\mu\nu}}{M_4^2},
\frac{\alpha k^2\tilde{S}^T_{\mu\nu}}{M_5^3}\bigg)\bigg] ~.  
\end{eqnarray}
The unspecified part of order $\alpha$ denotes       
the correctional terms by Eq.~(\ref{new}), 
which are negligible at low energies.  
With Eqs.~(\ref{soltrace}) and (\ref{solless}), 
we can find the full expression of $\tilde{h}_{\mu\nu}$,  
\begin{eqnarray} \label{final}
&&~~~\tilde{h}_{\mu\nu}(k,z=0)=\tilde{h}^T_{\mu\nu}(k,z=0)
+\frac{1}{4}\eta_{\mu\nu}\tilde{h}^0(k) 
\nonumber \\
&&=-\frac{2}{M_4^2k^2}\bigg[
\tilde{T}^{1}_{\mu\nu}(k)-\frac{1}{2}\eta_{\mu\nu}\tilde{T}^1(k)\bigg]
+O\bigg(\frac{\alpha\tilde{T}^{1T}_{\mu\nu}}{M_4^4},
\frac{\alpha \tilde{S}^T_{\mu\nu}}{M_5^3M_4^2}\bigg) ~.   
\end{eqnarray} 
Hence, at low energies, $\tilde{h}_{\mu\nu}$  
reproduces the graviton tensor structure 
that the 4D Einstein gravity theory predicts.     
Note that Eq.~(\ref{final}) is valid only when $\alpha\neq 0$ 
even if the effect of the $\alpha$ term is negligible at low energies. 
If $\alpha=0$, Eq.~(\ref{uv}) is no longer valid  
and different classes of $\tilde{h}_{\mu\nu}(k,z)|_{z=0}$'s solutions  
inconsistent with observed low energy 
gravity phenomena would also be allowed.  
With $T^1_{00}(x)=\rho(x)>>T^1_{ii}(x)$ ($i=1,2,3$),  
the non-relativistic gravity potential on the brane is 
given by~\cite{dvali}
\begin{eqnarray}
V(r)&=&\int dt\int\frac{d^4k}{(2\pi)^4}e^{-ikx}
\frac{1}{2}\tilde{h}_{00}(k,z=0) 
\nonumber \\
&\approx& -G_N\int d^3\vec{r}'
\frac{\rho(\vec{r}')}{|\vec{r}-\vec{r}'|} 
\end{eqnarray}
where $r\equiv \sqrt{(\vec{x}\cdot \vec{x})}$ and $G_N\equiv 1/(8\pi M_4^2)$.  
Therefore, the Newtonian $-\frac{1}{r}$ potential is guaranteed 
at {\it long} distances.  

\section{Conclusion}

In conclusion, only with  the conventional 5D gravity and scalar kinetic
terms, we have shown that 4D flat solution is always possible
independent of the magnitude of 4D brane cosmological constant
in 5D non-compact space-time with a $Z_2$ symmetry.
Since the ordinary SM fields are regarded as being localized at the brane,
4D vacuum energy by them and its variation do not destroy the 4D flatness.
To obtain the observed 4D gravity with non-compact extra space,
we introduce an additional brane-localized gravity kinetic term,
which successfully gives rise to $-\frac{1}{r}$ gravity potential 
on the brane.  
There are three ingredients for our solution: (i) The massless scalar 
is necessary toward a self-tuning solution,
(ii) the extra space should
be non-compact with a warp factor(the RS-II type model) so that
any fine-tuning condition is not needed, and
(iii) a general form of brane-localized gravity kinetic term is necessary
to realize the low energy Einstein gravity on the brane.  
While the self-tuning mechanism guarantees the background metric
of $\beta^2\eta_{\mu\nu}$,
the $-\frac1r$ potential is explained by the metric fluctuation $h_{\mu\nu}$.

\vskip 0.3cm

\noindent{\bf Acknowledgments}

\noindent
This work is supported in part by the BK21 program of Ministry of
Education(JEK), the KOSEF Sundo Grant(JEK), Korea Research
Foundation Grant No. KRF-PBRG-2002-070-C00022(JEK), and by DOE
under contract number DE-FG02-91ER40626(BK,QS).

\end{document}